\documentclass[aps,a4paper,twocolumn,showpacs]{revtex4-1}
\usepackage[xdvi]{graphicx}
\usepackage{latexsym}
\usepackage{amsbsy}

\newcommand{\om}{{ \mbox {\boldmath  $\omega$} }}
\newcommand{\Om}{{ \mbox {\boldmath  $\Omega$} }}
\newcommand{\fettmu}{{ \mbox {\boldmath  $\mu$} }}

\usepackage{epstopdf}

\begin{document}

\date{\today}

\title{Dynamics of magnetic single domain particles embedded in a viscous liquid} 

\author{K.\ D.\ Usadel$^1$} 
\author{C.\  Usadel$^2$}

\affiliation{ $^1$ Theoretische Physik,
  Universit\"{a}t Duisburg-Essen, 47048 Duisburg, Germany}
  
\affiliation{ $^2$ Inst. Biochem.,
  Universit\"{a}t M\"{u}nster, 48149 M\"{u}nster, Germany}

\begin{abstract}
Kinetic equations for magnetic nano particles dispersed in a viscous liquid are developed and analyzed numerically. Depending on the amplitude of an applied oscillatory magnetic field the particles orient their time averaged anisotropy axis perpendicular to the applied field for low magnetic field amplitudes and nearly parallel to the direction of the field for high amplitudes. The transition between these regions takes place in a narrow field interval.
In the low field region the magnetic moment is locked to some crystal axis and the energy absorption in an oscillatory driving field is dominated by viscous losses associated with particle rotation in the liquid. In the opposite limit the magnetic moment rotates within the particle while its easy axis being nearly parallel to the external field direction  oscillates.  The kinetic equations are generalized  to include thermal fluctuations. This leads to a significant increase of the power absorption in the low and intermediate field region with a pronounced absorption peak as function of particle size. In the high field region, on the other hand, the inclusion of thermal fluctuations reduces the power absorption. The illustrative numerical calculations presented are performed for magnetic parameters typical for iron oxide.  

  \end{abstract}
 \maketitle

\section{Introduction} %%%%%%%%%%%%%%%%%%%%%%%%%%%%%%%%%%%%
\label{s:intro}
In recent years magnetic nano particles are discussed intensively  in connection with biomedical applications,
in particular magnetic hyperthermia. The use of these particles in hyperthermia is based on the fact that power is absorbed in an applied oscillating magnetic field. The goal is to achieve a high absorption rate with particle parameters suitable for biomedical applications. This restricts the amplitude of the applied magnetic field to $100 - 200$ Oe and its frequency to $50 -1000$  kHz  \cite{usov_0,li,pank}.

Theoretically the dynamics of magnetic nano particles has been studied in recent years in very many papers 
for the case that the particles are fixed in space.  
The dynamics is then reduced to that of  magnetic moments in external fields for which the
stochastic Landau-Lifshitz-Gilbert (LLG) equation introduced by Brown
\cite{brown} is often used. This approach has gained increased interest recently  because
of its application to single-domain particles. In these nano particles (NPs) the internal degrees of freedom can be neglected at low temperatures 
making it possible to describe the whole particle as one macro spin of
constant length \cite{Ohand00}. Its dynamics is expected to be well described by a classical
approach. 
The dynamics of these macro spins is greatly influenced by thermal fluctuations originated from coupling of the spins to the surrounding medium (the heat bath). The theory of thermal fluctuations of small magnetic particles  
is one of the fundamental issues of modern magnetism \cite{hille}. 

As far as magnetic hyperthermia is concerned a recent comprehensive study is available which focusses on the problem of obtaining NPs with optimal parameters with respect to high energy absorption rates needed in biomedical applications \cite{carr}. However, the theoretical approach used in this work relies on various simplifications like linear response theory, two level approximation for magnetic NPs with uniaxial anisotropy or the concept of a frequency dependent coercivity. A more fundamental study based on the LLG equations has also been published recently \cite{usov_0}  which avoids the approximations mentioned. In both of these works the NPs are assumed to be fixed in space.  
  
In biomedical applications, however, a new aspect should be considered, the possibility for magnetic particles to move in the viscous liquid they are immersed in. The behavior of such particles has been studied in some detail, in particular in the theory of magnetic fluids \cite{ shliomis_0, coffey, shliomis, rosen}. 
In these systems the magnetic moment may rotate within the particle with respect to some crystal axis -i.e. Ne\'{e}l relaxation - and rotate together with the particle with respect to the liquid. Both processes  contribute to the power absorption of NPs when placed in a time varying magnetic field. Theoretical studies have assumed for the most part \cite{shliomis, shliomis_0} that both processes occur separately leading to the concept of effective relaxation times treated within  linear response theory \cite{rosen, lahonian,nedelcu}. This  heuristic approach  was critically discussed in detail in \cite{usov} because it certainly oversimplifies the physical situation, since it does not take into account the rich dynamics of magnetic NPs. Moreover the field amplitudes interesting for applications are beyond the linear response regime making a detailed investigation of the dynamics of mobile NPs in finite alternating fields necessary. 

Therefore, in the present paper this dynamics is studied starting from the well established LLG equations for magnetic NPs suitable modified for taking the rotational degrees of freedom into account. Our aim is to establish a set of kinetic equations for studying the dynamics of mobile single domain NPs at finite temperatures.   A similar approach has been discussed and applied to magnetic hyperthermia very recently in \cite{mam,usov}. In these papers, however, sets of kinetic equations are proposed which both violate in limiting cases basic conservation laws to be valid on physical grounds making the results obtained doubtful.  An important achievement of the present investigation is the proposal of a set of  kinetic equations which are in accordance with necessary conservation laws. 

The NPs to be considered are assumed to have an uniaxial anisotropy characterized by an unit vector $\bf n$ parallel to the easy axis. Their magnetic moment has a fixed absolute value $\mu_s$ and is oriented parallel to the unit vector $\bf e$.  Numerical solutions of the kinetic equations proposed  show that depending on the amplitude of an applied oscillating magnetic field basically two different regions occur in the stationary state. In a low field region    $\bf n$  and $\bf e$ 
 are nearly parallel and their time average is perpendicular to the direction of the applied field. In the high field region, on the other hand,   $\bf n$ is nearly parallel to the direction of the applied field while $\bf e$ oscillated between the states $\pm \bf z$ where $\bf z$ is the direction of the applied field. Such a scenario has  been found and discussed already in \cite{usov}. The results obtained, however,  differ from our findings  in many details. In particular, we found an extremely sharp transition between these regions at an applied field amplitude of about one half of the coercive field accompanied by a dramatic increase of the area of the hysteresis loop when passing from the low field to the high field region.
  
Thermal fluctuations play an important role in the dynamics of magnetic NPs. The kinetic equations proposed are generalized accordingly. Numerical solutions of these equations are focused  on a calculation of the power absorbed by the system, particularly on its dependence on the size of the particles.  For low field amplitudes a resonance type absorption peak as function of particle size is found at finite temperatures. The corresponding power absorption is much higher than without thermal fluctuations. This enlarged power absorption originates from temperature driven switching processes of the magnetization relative to the NPs and therefore is strongly dependent on the volume of the NPs.  For increasing field amplitudes these absorption peaks first broaden and then go over to a monotonous behavior in the high field region. In this region thermal fluctuation reduce the power absorption especially for very small particle sizes. Results are compared with those obtained for an ensemble of NPs with fixed randomly distributed anisotropy axis. In this case the power absorption is generally smaller than in the case of mobile particles.  In view of applications in hyperthermia the illustrative numerical calculations  are performed for magnetic parameters typical for iron oxides at room temperature.

\section{Basic Equations}

A nano particle  considered in the present paper is modeled as a uniformly magnetized spherical rigid body (RB) with uniaxial magnetic anisotropy so that its Hamiltonian is given by     
\begin{equation}
{\cal H} = 1/2  \, {\Theta  } \, { \mbox {\boldmath  $\omega$} }^2 -  { \mbox {\boldmath  $\mu$} } \cdot {\bf B}  - \tilde D ( { \mbox {\boldmath  $\mu$}} \cdot \bf n )^2 .
\label{e:hami}
\end{equation}
$\Theta$ denotes the moment of inertia of the RB, $\om$ its angular velocity, $\fettmu$ the magnetic moment with magnitude $\mu_s$, $\tilde D {\mu_s}^2$ an anisotropy energy,  $\bf B$ the external magnetic field and $\bf n$ an unit vector parallel to the anisotropy axis of the RB. This vector often called director is firmly attached to the RB so that its equation of motion is  given by 
 \begin{equation}
\frac{\partial {\bf n}}{\partial t} =  \om \times \bf n .
\label{e:n_punkt}
\end{equation}
The differentiation in Eq. (\ref{e:n_punkt}) is performed in a coordinate system fixed in space, the laboratory frame.

The equations of motion for the angular momentum   ${\bf L} = \Theta \, \om$ of the RB and for the spin momentum $\bf S$ associated with the magnetic moment $\fettmu$ are determined by the corresponding torques given by  
\begin{equation}
\frac{\partial {\bf L}}{\partial t} = {\bf M}_l
\end{equation}
and
\begin{equation}
\frac{\partial {\bf S}}{\partial t} = {\bf M}_{s}.
\label{e:s_punkt_1}
\end{equation}
The torque ${\bf M}_l$  can be obtained  from the energy change following an infinitesimal 
rotation of the anisotropy axis $\bf n$,     $\delta \bf n =\delta { \mbox {\boldmath  $\varphi$} } \times \bf n $, according to $ {\bf M}_{l} \cdot \delta { \mbox {\boldmath  $\varphi $} }  = - \delta \cal H$  with the result that \cite{usov}
\begin{equation}
{\bf M}_{l} =  2\, \tilde D (\fettmu \cdot \bf n ) \, \bf n \times \fettmu .
\label{e:M_r}
\end{equation}
The torque  $\bf M_{s}$ can be obtained in a similar way from the energy change following an infinitesimal rotation of the moment,    $\delta \fettmu =\delta { \mbox {\boldmath  $\varphi$} } \times \fettmu $,  %,  according to 
%$ {\bf M}_{s} \cdot \delta { \mbox {\boldmath  $\varphi $} }  = - \delta \cal H$ 
with the result that
\begin{equation}
{\bf M}_{s} = \fettmu \times {\bf B} + 2\, \tilde D (\fettmu \cdot \bf n ) \, \fettmu \times \bf n .
\end{equation}
The sum of both is the net torque %${\bf M}_l + {\bf M}_s = \fettmu \times \bf B$ 
in which the intrinsic contributions stemming from the anisotropy energy cancel each other. 
The equation of motion for  the total angular momentum  $\bf S +{ \bf L}$  is thus given by
\begin{equation}
\frac{\partial {(\bf S + \bf L)}}{\partial t} = \fettmu \times {\bf B }
%\label{e:s_punkt}
\end{equation}
showing that it is conserved in zero field as it should be. Note that angular momentum $\bf L$ and spin momentum $\bf S$ are coupled through the anisotropy energy $ - \tilde D ( { \mbox {\boldmath  $\mu$}} \cdot \bf n )^2 $ so that the conservation of the total angular momentum is nontrivial. 

So far we have only considered the conservative part of the equations of motion. Dissipative contributions are added phenomenologically. Here we have to distinguish between internal and external relaxation processes. External processes arise due to the interaction of the NPs with the surrounding fluid in the form of rotational damping and thermal fluctuations. 

Internal relaxation processes to be considered first occur due to the motion of the spin relative to the NP. For this spin dynamics we introduce a Gilbert damping term proportional to $\bf S \times \frac {\partial {\bf  S} }{\partial t}$. Because it arises from the motion of the spin $\bf S$  it must vanish if this vector does not move relative to the RB. Therefore in the present case the time derivative in this term must be interpreted as time derivative in the moving frame fixed to the RB. Thus in terms of derivatives in the laboratory frame the Gilbert damping term should be written as  -$\tilde \alpha \,(\bf S  \times (\frac {\partial {\bf  S} }{\partial t} - \om \times \bf S ))$ where $\tilde \alpha$ denotes a damping constant. This term has to be added to the right hand side  of Eq.\ (\ref{e:s_punkt_1}) which together with Eq. (6) can be rewritten as
 \begin{equation}
\frac{\partial {\bf S}}{\partial t} = \fettmu \times {\bf B}_{e} -\tilde \alpha (\bf S \times (\frac {\partial {\bf  S} }{\partial t}-\om \times \bf S) )
\label{e:s_punkt}
\end{equation}
where ${\bf B}_e$ denotes an effective field, 
\begin{equation}
{\bf B}_{e}= {\bf B}+ 2\, \tilde D (\fettmu \cdot \bf n ) \,  \bf n .
\end{equation}

The damping term discussed is a torque acting on the spin moment due to the interaction of the magnetic moment with the other degrees of freedom of the RB. Therefore, the opposite momentum is transferred to the rotational degrees of freedom, i.e. the Gilbert damping term has to be subtracted from the right hand side of Eq. (5)  resulting in
\begin{equation}
\frac{\partial {\bf L}}{\partial t} = - 2\, \tilde D (\fettmu \cdot \bf n ) \, \fettmu \times \bf n +\tilde \alpha (\bf S \times (\frac {\partial {\bf  S} }{\partial t}-\om \times \bf S) ). 
\label{e:l_punkt}
\end{equation}
It follows that  Eq. (7) is still valid so that for an isolated particle the total angular momentum $\bf L + \bf S$ again is conserved in zero field as it should be irrespectively of internal damping processes. 
The results obtained  are in contrast to Ref. \cite{usov} in which the second term on the right hand side of Eq. (10) is missing.  They also differ from equations proposed in \cite{mam} which were already criticized of being incorrect in \cite{usov}.  

Further sources of damping are external relaxation processes stemming from  the rotational motion of the NP in the surrounding liquid.  We assume a damping term  linear in $\om$  \begin{equation}
{\bf N}_l = - \xi \om
\end{equation}
with friction coefficient $\xi = 6 \eta V_{d}$. Here, $\eta$ is the dynamical viscosity while $V_{d}$ denotes the hydrodynamic or total volume of the particle, i.e. it is assumed that the particle  consists of a magnetic core region of volume $V_m$ with radius $r_m$ eventually covered by a nonmagnetic surfactant layer.  Here it is assumed that the surfactant layer is firmly attached to the magnetic core. 

The torque $\bf N_l$ has to be added to the right hand site of Eq. (10) so that the equation of motion for $\bf L$ now reads
\begin{equation}
\frac{\partial {\bf L}}{\partial t} = - 2\, \tilde D (\fettmu \cdot \bf n ) \, \fettmu \times \bf n +\tilde \alpha (\bf S \times (\frac {\partial {\bf  S} }{\partial t}-\om \times \bf S) )   - \xi \om.
\label{e:l_punkt}
\end{equation}
The equation of motion for the total angular momentum  $\bf S +{ \bf L}$  is therefore given by
\begin{equation}
\frac{\partial {(\bf S + \bf L)}}{\partial t} = \fettmu \times {\bf B } + {\bf N}_l.
\label{e:sl_punkt}
\end{equation}

 To proceed we first solve Eq. (8) for $\frac{\partial {\bf S}}{\partial t} $ with the result 
\begin{eqnarray}
\frac{\partial {\bf S}}{\partial t} = \frac {1}{1+ {\tilde \alpha}^2 \, \bf S \cdot \bf S } \Big (\fettmu \times {\mathbf B}_{e} +\tilde \alpha \bf S \times (\om \times \bf S)  \nonumber \\-\tilde \alpha (\bf S \times (\fettmu \times { \mathbf B }_{e} )+ \tilde \alpha {\bf S}^2 \, \bf S \times \om) \Big ).
\end{eqnarray}
An explicit  equation of motion for $\bf L$ can then be obtained by inserting Eq. (14)  into Eq. (13).

Finally we introduce an unit vector $\bf e$ in the direction of the magnetic moment, $ \bf e = \fettmu / \mu_s$ .The spin momentum $\bf S $ can be expressed as $\fettmu =  - \gamma \, \bf S $ where $\gamma$ denotes the gyromagnetic ratio. We express $\bf S$ and $\fettmu$ in terms of $\bf e$, define $D = {\mu_s}^2  \tilde { D} $, $ \alpha = \tilde \alpha \, \sqrt {{\bf S } \cdot {\bf S}}$ and end up with the following set of equations
\begin{eqnarray}
\frac{\partial {\bf e}}{\partial t} = - \frac {\gamma}{1+ {\alpha}^2  } \Big (\bf e \times {\mathbf B}_{e} + \alpha \,\bf e\times (\bf e \times { \mathbf B }_{e} ) +\nonumber \\  \frac {\alpha} { \gamma }\, \bf e \times (\om \times \bf e) + \frac {\alpha ^2}  {\gamma} \bf e \times \om \Big)
\end{eqnarray}

\begin{eqnarray}
 \Theta \, \frac{\partial {\om}}{\partial t}= \frac {\mu_s}{\gamma} \frac{\partial \bf e}{\partial t} + \mu_s \bf e \times \bf B - \xi \, \om.
\label{e:omega_punkt_2}
\end{eqnarray}
These equations together with Eq.\ (\ref{e:n_punkt})  are the basis of the following calculations. 
At low temperatures they specify the dynamics of the NP completely. At elevated temperatures thermal fluctuations have to be added to these  equations. This will be discussed in section IV.

 Note that the quantity $\alpha$ appearing in Eq. (15) is the dimensionless damping parameter usually used in the literature \cite{garcia, nowak, usad_1}.  In the present paper the kinetic equations proposed will be discussed exclusively in the low frequency limit because only this limit is relevant for biomedical applications. Note, however, that the equations proposed are not restricted to this limit.

\section{Application to a NP in a viscous liquid}

A cartesian coordinate system spanned by unit vectors  ${\hat {\bf x}}$, ${\hat {\bf y}}$ and ${\hat {\bf z}}$ and fixed in space is used.
The applied  alternating magnetic field is assumed to be aligned parallel to one of these axis, the $z$-axis, and is given by 
  \begin{equation}
 {\bf B}  = B_0 \,{\mathrm {cos}}( \tilde \omega \,t) \,{\hat {\bf z}}
 \label{e:t}
 \end{equation}
 where $\tilde \omega = 2 \pi f$ is the angular 
 frequency of the external driving field.   For practical applications in hyperthermia the frequency $f$ is of the order of $50 - 1000 $ \,kHz. 
For the numerical calculations  it is convenient to use the 
dimensionless field amplitude  $b_0  = B_0 / B_c$ where $B_c = 2D/\mu_s$  denotes the coercive field.  We scale the time $t$ according to $\tau = t \, \omega_l$ 
 where  $\omega_l = 2  \gamma D /\mu_s$
 is the Larmor frequency and we introduce  the reduced angular velocity $\Om =  \om / \omega_l $. The resulting dimensionless equations are easily  seen  to depend only on the parameters  $R=\Theta {\omega_l}^2/\mu_s B_c $,  $\Xi = 6 \, \eta \, \gamma V_{dyn} / {\mu}_s$ and $F = 2\pi f/\omega_l$. 
 
To perform illustrative numerical calculations we choose magnetic parameters typical of the particles of iron oxides (magnetite) \cite{usov}. The anisotropy energy $D$ can be expressed as $D=K_1 \cdot V_m$ and $K_1$ is the magnetic anisotropy constant given by  $K_1 = 10^4\,J/m^3$ for magnetite. The magnetic moment ${\mu}_s$ can be expressed as ${\mu}_s = M_s \cdot V_m$ with $M_s=4 \cdot 10^5 A/m$. With these values one obtains  for the coercive field $B_c = 50$ mT and for the reduced friction coefficient $\Xi = 2.64  \cdot 10^3 \, \cdot \eta / \eta_{water} \, \cdot V_{d}/V_m \, $ with $\eta_{water} = 10^{-3} \cdot $ kg/ms.

In the following we use 
\begin{equation}
\tilde \eta = \eta / \eta_{water} \, \cdot V_{d}/V_m 
\label{e:eta_tilde}
\end{equation}
because the dynamical viscosity $\eta$ and the hydrodynamic volume $V_d$  enter in this combination throughout.  Note that $\tilde \eta $ is the sole parameter depending on the interaction of the NP with the surrounding fluid in the absence of thermal fluctuations.

For the parameters underlying our numerical calculations the kinetic equations  can be simplified if one takes into account \cite{usov, newman} that because of the presence of a viscous damping together with the smallness of the NP it is possible to neglect the term involving the moment of inertia, assuming $R=0$, so that $\om$ can explicitly be obtained from Eq.\ (\ref{e:omega_punkt_2}). To study the accuracy of this approach we performed calculations with and without using this approximation and found no detectable  differences in the results.   
 
The kinetic equations are solved numerically.  Initial conditions for $\bf e$, $\bf n$ and $\Om$
 at  time $t=0$  have to be specified. Usually the reduced magnetization is chosen to be ${\bf e} = \hat{\bf x}$ while the easy axis of the NP is set parallel to  $  \hat {\bf x} + \hat {\bf z} $ and $\Om = 0$.  The system of equations is damped so that a stationary state is reached finally.

Under the influence of an oscillating  driving field the magnetization switches initially on very small time scales of the order of $\tau_{init} \sim 10 - 20$ 
towards the direction of the easy axis into a state parallel to the effective field ${\bf B}_e$. Note that on these short time scales the applied field is practically constant. After reaching this position magnetization and anisotropy axis start oscillating with the frequency of the applied field accompanied by a slow relaxation into a stationary state. 
  This relaxation process is  shown for two values of the amplitude of the driving field in Fig. 1. 
 \begin{figure}[h] 
   \centering
 \includegraphics[width=3.5in]{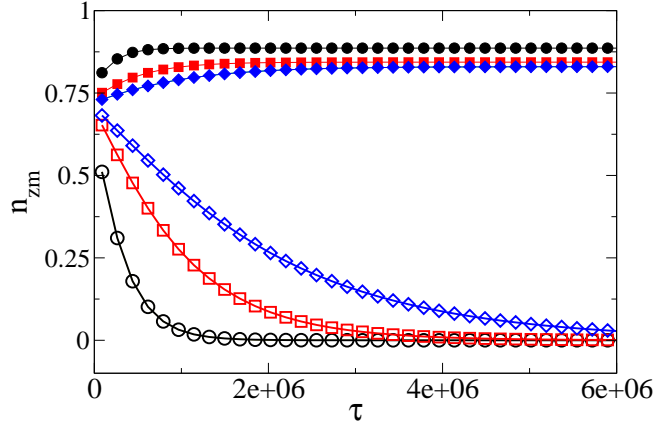}  \\[-0.1cm]
   \caption{(color online) Averaged $z$-component of the anisotropy axis for $b_0=0.54$ (upper three curves) and $b_0=0.48$ (lower three curves).   $\tilde \eta   = 10$ (circles, black),  $30$ (squares, red) and $60$ (diamonds, blue). $f=50$ kHz, $\alpha = 0.5$.}      
   \label{f:fig_1}
   \end{figure}  
Here, ${ n}_{z m}$ denotes averaged values of the  $z$-component of the anisotropy axis  averaged over periods of the applied field  for $f=  50$ kHz.  The symbols in Fig. 1 represent these quantities ${ n}_{z m}$  which are set at odd multiples of ${\tau}_0/2$ where ${\tau}_0$ denotes the period of the driving field which in the present case is given by $\tau_0=1.76 \,\,\cdot10^5$, i. e. huge as compared to $\tau_{init}$. Closed symbols refer to $b_0=0.54$  while open symbols refer to the smaller field $b_0=0.48$. Results for three  different values of $\tilde \eta$ are shown:  circles refer to  $ \tilde \eta = 10$, squares refer to  $\tilde \eta = 30$ while diamonds refer to $\tilde \eta = 60$. 

Fig. 1 shows that the averaged component $n_{zm}$  decays  to zero for $b_0 = 0.48$ which means that  in the average $\bf n$ orients perpendicular to the driving field in the stationary state. On the other hand for  $b_0=0.54$ in the average $\bf n$ orients nearly parallel  to the direction of the driving field.  An increase in the viscosity only slows down this relaxation into the stationary states.    
 
The results discussed are typical in the sense that independently of details in the parameters two states are found, a state for small fields in which the direction of the anisotropy axis averaged over field periods is orientated perpendicular to the applied field and a state for larger fields in which it is oriented nearly parallel to the field. 
We now discuss more details of the dynamics in the stationary states. 
\begin{figure}[t] %  figure placement: here, top, bottom, or page
   \centering
 \includegraphics[width=3.5in]{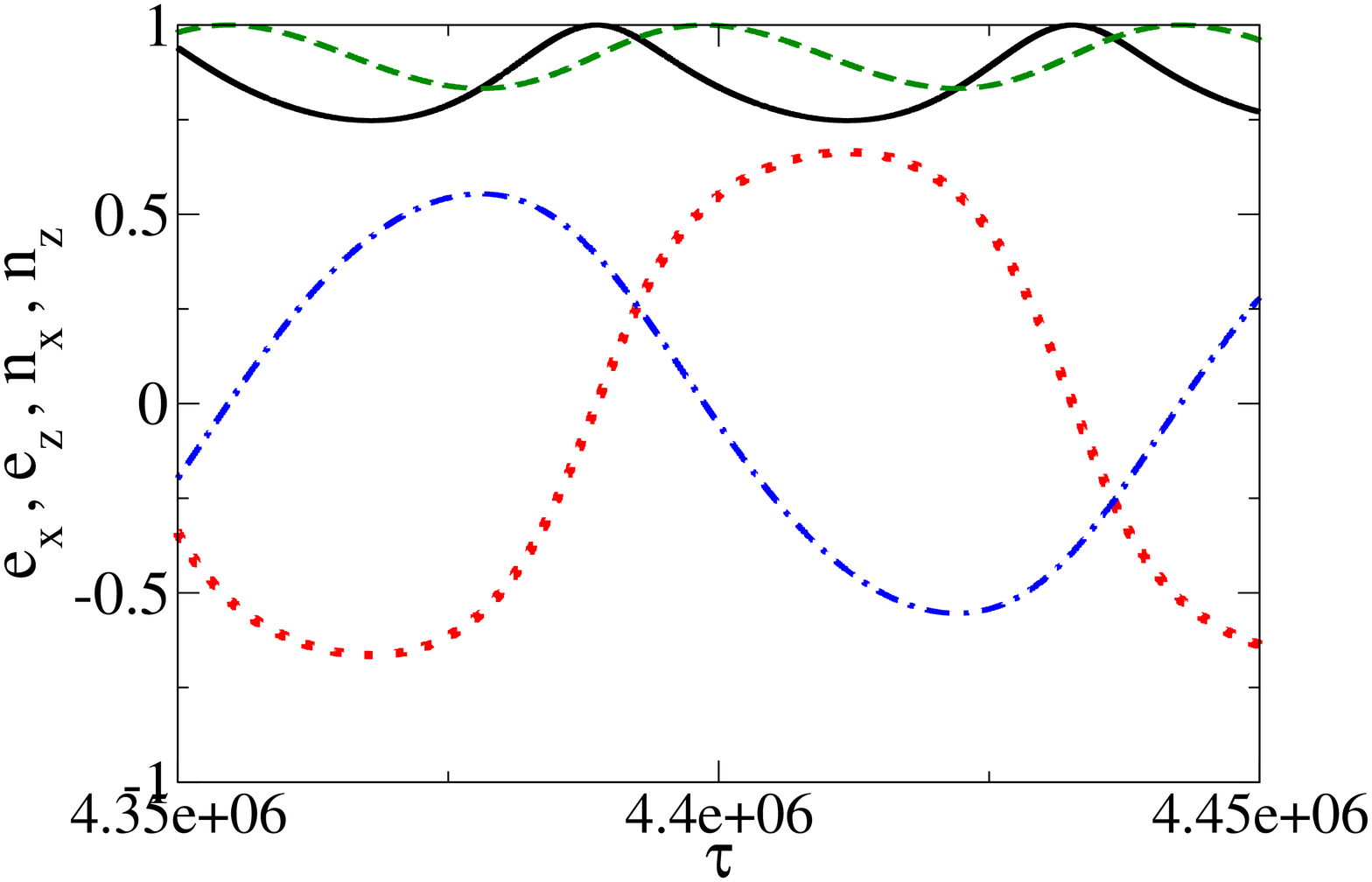}  \\[-0.1cm]
  \includegraphics[width=3.5in]{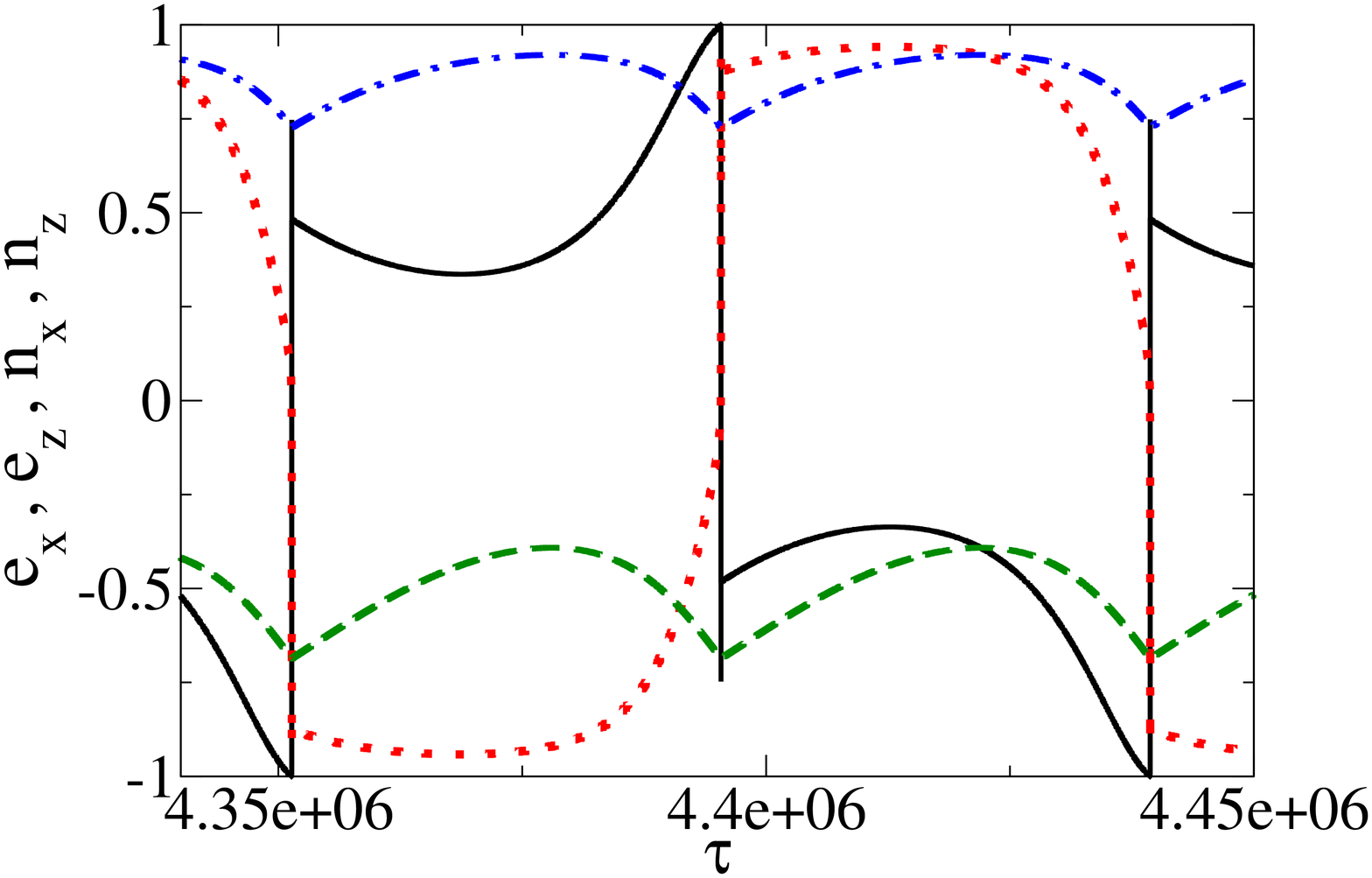}  \\[-0.1cm]
   \caption{(color online) $e_x$ (solid black),  $e_z$ (dotted red),  $n_x$ (dashed green), $n_z$ (dotted-dashed blue).
    $\alpha = 0.5$, $\tilde \eta  = 10$, \\$f=50$ kHz. 
   Upper panel: $b_0=0.52$, lower panel $b_0=0.53$.}      
   \label{f:fig_2}
   \end{figure}

In Fig. \ref{f:fig_2} components of $\bf e$ and $\bf n$ are shown in the stationary state as function of reduced time $\tau$ for field amplitude $b_0 = 0.52$ in the upper panel and for $b_0=0.53$ in the lower panel. Shown are the amplitudes after about $20$ field cycles sufficient for reaching stationarity (for $\tilde \eta = 10$). For $b_0=0.52$ the x-components of both the anisotropy axis, $n_x$ (dashed green), and the reduced magnetization, $e_x$ (solid black), are close to one with small  oscillations in time.  As was discussed above in this case the NP is orientated in such a way that its easy axis is on the average parallel to the x-axis, i.e. perpendicular to the driving field.  The magnetization is practically locked to  the NP. This can be seen even more clearly by considering the scalar product $\bf e \cdot \bf n$ (not shown) between the scaled magnetization and the director of the NP which in the present case turns out to deviate from one by less than 5 \%. The  components $n_z$ (dotted dashed blue) and $e_z$ (dotted red), however, both  oscillate around zero while the corresponding y-components are extremely small. Thus, with numerical accuracy $\bf n$ lies in the $x-z$ plane with a large component parallel to the $x$-axis and a smaller $z$-component which oscillates  around zero. Therefore the angular velocity $\Om$ (not shown) is practically parallel to the $y-$direction. In the present case its $y$-component turns out to be  of the order of $10^{-4}$  while the other components of $\Om$ are negligible.

For larger fields a completely different behavior is observed. The lower panel in Fig. \ref{f:fig_2}  shows that for $b_0=0.53$  the z-component of $\bf n$  (dashed dotted blue) is nearly aligned with the direction of the driving field (the $z$-axis) which means that the particle has rotated into a direction with the easy axis nearly parallel to the driving field. The magnetization $\bf e$, on the other hand, rotates between  the states ${ \pm \bf \hat z} $ in the $x-z$-plane.  %In this case  friction and therefore energy dissipation arises due to an oscillation of the magnetization relative to the NP (N\'eel relaxation) while in the first case friction is mainly due to oscillations of the NP relative to the surrounding fluid (Debye relaxation) with a magnetization practically fixed relative to the NP. 
It is important to note that in both cases the amplitude of the driving field is well below the coercive field $b_c=1.0$. 

\begin{figure}[t] 
   \centering
    \includegraphics[width=3.5in]{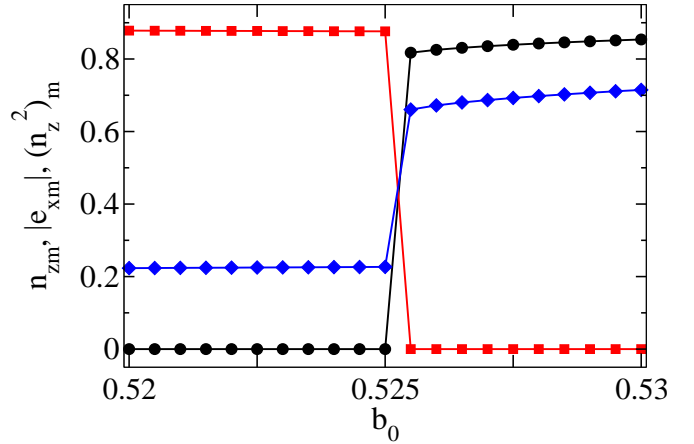}  \\[-0.1cm] 
     \caption{(color online) Field dependence of averaged components: $n_{zm} $ (dots, black), $e_{xm}$ (squares, red), $(n_z^2)_m$ (diamonds, blue). $\alpha = 0.5$, $f=50$ kHz, $\tilde \eta   = 10$.}      
      \label{f:fig_3}
   \end{figure}

The two cases discussed are typical in the sense that there exists a low field region in which $\bf e$ is locked within the RB with an easy axis oriented in the average perpendicular to the driving field and a high field region in which $\bf e$ rotates within the RB. For a quantitative analysis we calculated the time averaged x-component of the reduced magnetization in the stationary state, $e_{xm}$, as well as the time averaged z-component of $\bf n$ averaged over periods of the oscillating field as a function of the amplitude of the driving field.  From the discussion above we expect $e_{xm}$ close to one for small fields and zero in the large field region while the opposite behavior is expected for $n_z$. The results  shown in 
Fig. \ref{f:fig_3} confirm this. A remarkable sharp transition is observed between these regions at a threshold field $b_{th}$ around $0.525$. Also shown in Fig. \ref{f:fig_3} is $({n_z^2})_m$ which jumps from a small value below the threshold to a large value above it showing that in the low-field region there are finite oscillations of $ n_z$ around its mean value $n_{zm}=0.0$ while in the high field region these oscillations are around a finite mean value.
   
The switching  behavior of immobile NPs is frequently discussed within the Stoner - Wohlfarth model \cite{sw} in which the energetically stable and metastable positions of  the magnetization of a single domain NP with uniaxial anisotropy in the presence of a static magnetic field are considered. It turns out that the minimum static field needed to reverse  the magnetization  at zero temperature depends on the direction of the applied field with respect to the easy axis as described by the well-known  Stoner-Wohlfarth astroid. The smallest switching field is given by $B_c/2$ and therefore close to the threshold field found in the present case. Note, however, that a complete  agreement with this result is not expected since on the one hand time-dependent magnetic fields as considered  here reduce the required minimum field below the Stoner-Wohlfarth limit \cite{sun_lett, sun, usad_2} and on the other hand the mobile NPs are not forced into a specific direction with respect to the applied field. Nevertheless the resemblance of minimum switching field and threshold field seems to be remarkable.    

Additional calculations show that on the one hand side the clear separation between a low field region with $n_{zm} =0$ and a high field region with a finite value of $ n_{zm}$ is observed for all values of the parameters studied. On the other hand the precise value of the threshold field separating these regions although always found to be around $b_0 \sim 0.5$ turned out to depend on parameters like damping, viscosity or frequency. Thus the driven dynamical system can settle in various stationary states which become accessible by changing the system parameters. A detailed study of this complex dynamics is not followed any further here because we are mainly interested in the finite temperature behavior of the system for which these details are expected to be not so important. 
%Additional calculations showed that this field is slightly affected by parameters like frequency, particle size or viscosity.
 \begin{figure}[t] %  figure placement: here, top, bottom, or page
   \centering  
   \includegraphics[width=3.5in]{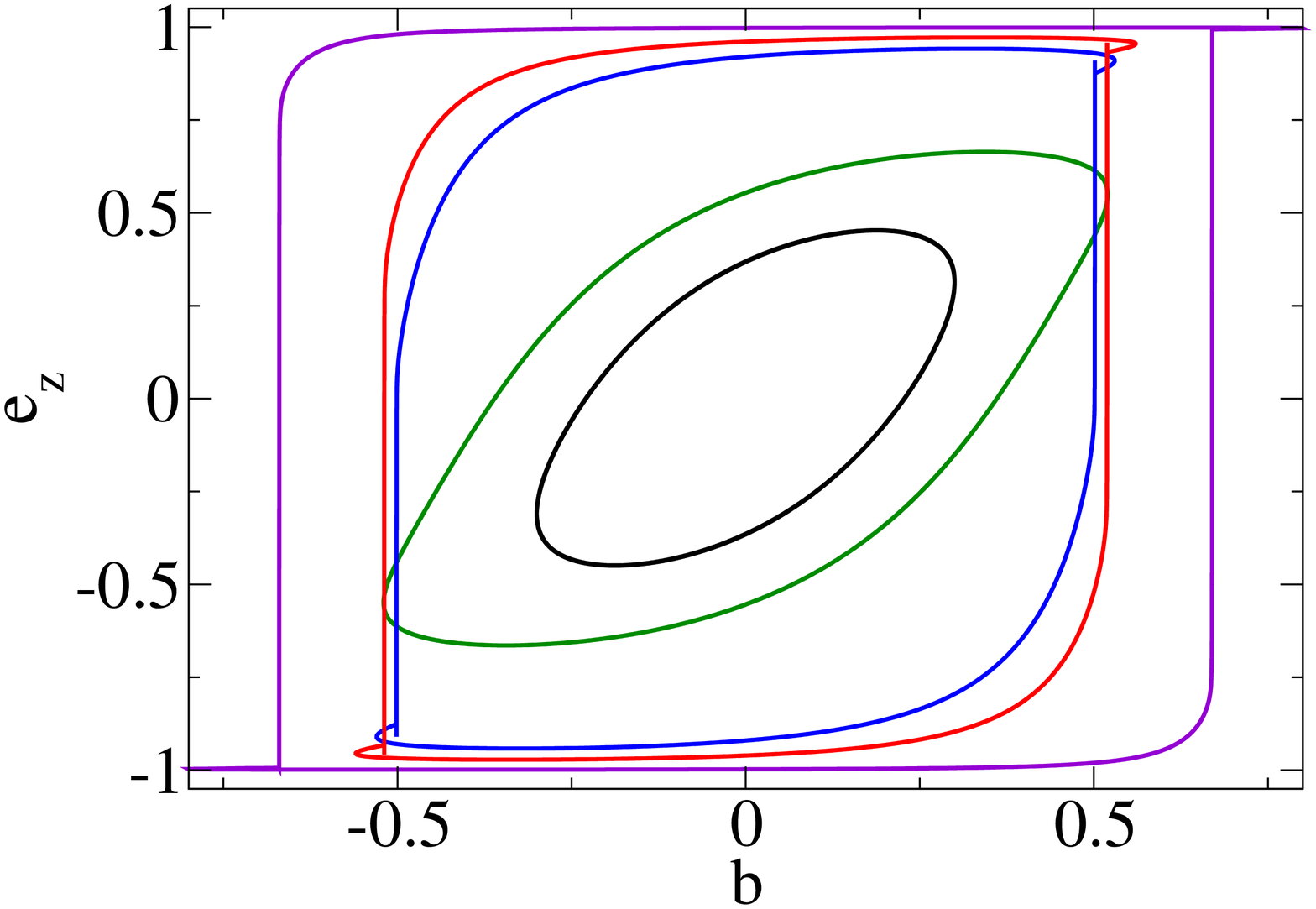}  \\[-0.1cm]
    \includegraphics[width=3.5in]{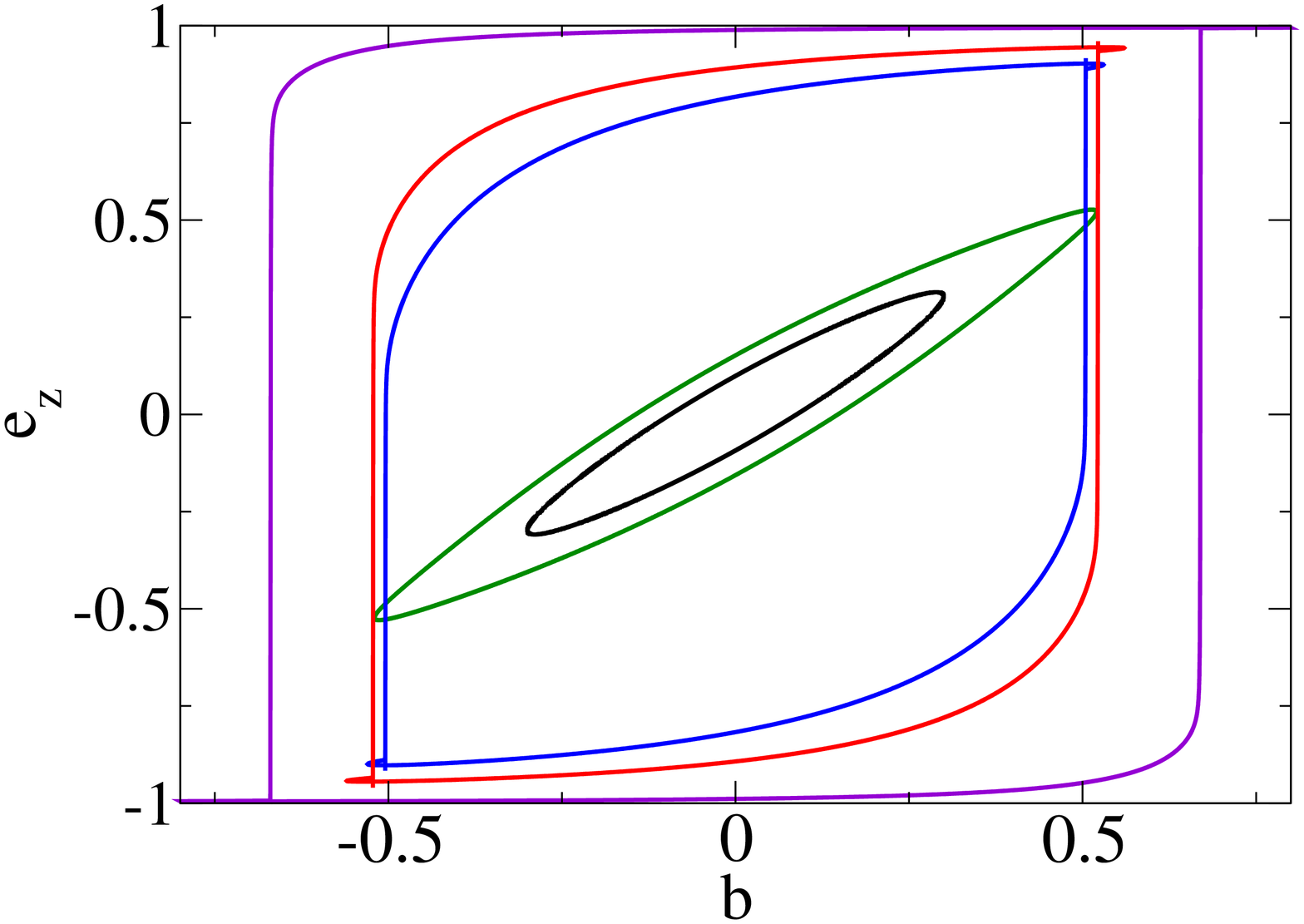}  \\[-0.1cm]
   \caption{(color online)  Hysteresis loops for $f=50$ kHz (upper panel) and for $f=200$ kHz (lower panel). With increasing order of the loop area field values are $b_0 = 0.3, 0,52, 0,53, 0.56, 0.8$.}      
   \label{f:fig_4}
   \end{figure}  
   
The transition between the two regions discussed above is accompanied by a dramatic change in form and area of the hysteresis loops. Typical loops are shown in Fig. 4 for low frequencies, $f=50$ kHz (upper panel) and $f=200$ kHz (lower panel). The loop for $b_0=0.3$ (black) has an ellipsoidal shape with a small area. It widens slightly when going up with the field  to $b_0=0.52$ (green). Increasing the field further to $b=0.53$ a drastic increase of the loop area is observed together with a change  to a more rectangular shape. With increasing frequencies the loop area decreases significantly in the low field region  while in the high field region no dramatic changes are found as can be seen in  Fig. \ref{f:fig_4}.

The area of the hysteresis loops is proportional to the power absorbed. Our calculations show that a dramatic increase of the absorption within a narrow field interval occurs when passing from the low to the high field region around  $b_{0} \sim 0.5$. In the high field region the magnetization switches relative to the NP in contrast to the low field region where only oscillations of the NP relative to the viscous fluid occur.  Thus the large increase of the energy absorption observed  originates obviously from the transition of oscillatory dynamics to switching of the magnetization relative to the NP. We note in passing that due to the reduced squareness of the hysteresis loop an estimate of its area only from data of the dynamical coercivity as was done in Ref. \cite{carr} is not meaningful for small  field amplitudes.

The absorbed power can be obtained either by measuring the loop area  or alternatively from \cite{usov_0, rosen, usov, usad_1, carr}
\begin{equation}
{\cal P}= -{\mu}_s  \langle{\bf e}\frac {\partial {\bf B} }{\partial t}\rangle.  
\label{e:power1}
 \end{equation}
The use of this expression is more convenient  especially at finite temperature where large thermal fluctuations have to be taken care of.
The brackets in Eq.(\ref{e:power1}) denote a time average over full field cycles of the driving field in the stationary state. 
 In the present case the reduced averaged absorbed power 
%\begin{equation}
 $p_{av} = {\cal P}/2K_1 V_m$
%\end{equation}
can be expressed explicitly as 
\begin{equation}
{ p_{av}}= {\tilde \omega} \, b_0 \,  \langle { e_z sin(\tilde \omega t)} \rangle.  
\label{e:power2}
 \end{equation}
Note that the quantity  $\bf e$ and therefore  $p_{av}$ are independent of the volume of the NPs in the limit $R\rightarrow 0$ as a  closer inspection of Eqs. (15) and (16) shows.   
 \begin{figure}[t] 
   \centering
 \includegraphics[width=3.5in]{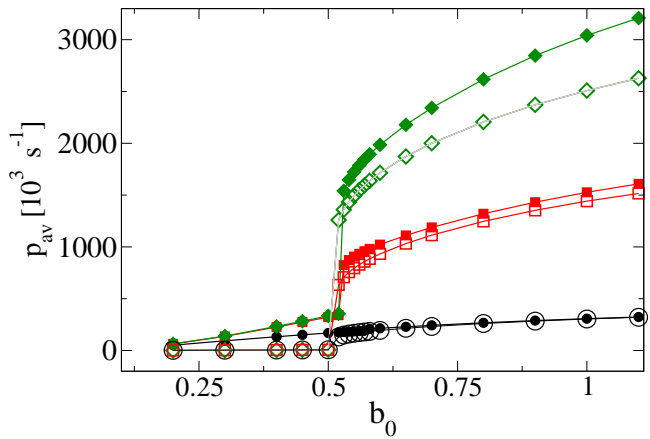}  \\[-0.1cm]
  \caption{(color online) 
 Averaged absorbed power $p_{av}$ vs.  $b_0$ for different values of $f$. Closed symbols are for $\tilde \eta =1.0$, open symbols for $\tilde \eta =50$. $f=1000$ kHz: diamonds, green; $f=500$ kHz: squares, red; $f=100$ kHz: circles, black. $\alpha =0.2$.}
 \label{f:fig5}
 \end{figure}  
 
In  Fig. \ref{f:fig5}  we show $p_{av}$   
with the same sets of parameters as used above  for different values of the driving frequency $f$. The solid symbols show results for a reduced viscosity $\tilde \eta = 1.0$  while open symbols represent results for $\tilde \eta =50$. The expected dramatic increase of the absorbed power when passing through the threshold region 
is clearly visible. 
Furthermore a strong increase with increasing frequency is observed. On the other hand, an increase of the effective viscosity $\tilde \eta$  leads to a significant  reduction of $p_{av}$ especially for high frequencies and in the low field region where  $p_{av}$ is dramatically reduced down to about $6 \cdot 10^3 \, s^{-1}$ for $\tilde \eta =50$ for all frequencies shown. The reason for this finding obviously is the reduced mobility of the NPs leading to less frictional losses. 

As was noted above  for a fixed value of the effective viscosity $\tilde \eta$ the reduced absorbed power $p_{av}$ is independent of the magnetic volume of the NPs. Thermal fluctuations, on the other hand,  are strongly volume dependent  leading therefore not only to quantitative but also to qualitative changes of the results presented. In particular, we will show in the next sections that a resonance type absorption peak as function of particle size occurs for small amplitudes of the driving field. It originates from temperature driven switching processes of the magnetization which have a strong influence on the power absorption.

 \section{Thermal fluctuations}
   
Thermal fluctuations are expected to have a significant impact on the behavior of NPs, especially if their  radius becomes very small. 
Here  we assume that the spin dynamics is governed by the stochastic
Landau-Lifshitz-Gilbert (LLG) equation \cite{brown} in which a fluctuating field $\boldsymbol \zeta$ is added to the effective field ${\bf B}_e$. This fluctuating field is generally assumed to be Gaussian distributed with zero mean and
correlator \cite{brown}
\begin{equation}
  \langle \zeta_{l}(0)\zeta_{m}(t) \rangle_\zeta =
   \delta_{l,m} \delta(t) 2 \alpha k_B T / (\mu_s \gamma).
  \label{e:corr4}
\end{equation}
In the same spirit the equation of motion for $\boldsymbol \omega$ has to be modified by adding a fluctuating torque $\boldsymbol \epsilon $ to the right hand site of Eq. (16) which leads to Brownian rotation of the RB in the liquid \cite{coffey, coffey_kal}.  $\boldsymbol \epsilon $ is also assumed to be Gaussian distributed with zero mean and with correlator \cite{coffey_kal}
\begin{equation}
  \langle \epsilon_{l}(0)\epsilon_{m}(t) \rangle_\zeta =  \delta_{l,m} \delta(t) 2 \, \xi \, k_B \,T .
  \label{e:corr5}
\end{equation}
In both of these equations  $l$ and $m$ label cartesian components of the fluctuating fields. 

In the following the stochastic equations are solved numerically with emphasis  on a calculation of the power absorption by a set of NPs.  Methods well known from literature are used \cite{garcia, nowak, usad_1}. The power absorption is calculated using  Eq. (\ref{e:power1}) in which the brackets now denote an average over a very large number of field cycles in order to reduce the effect of thermal fluctuations. Note that alternatively one can split this into an ensemble average over a set of NPs with a correspondingly reduced number of field cycles. In the following we average over about $4000$ field cycles altogether after reaching a stationary state.  
We found that this value is sufficient in general to obtain reproducible results.    

Thermal fluctuations depend strongly on the volume of the NP. It is therefore of interest to study the $r_m$ - dependence of $p_{av}$ in detail. 
A reduction  to dimensionless units reveals that for a fixed set of  magnetic parameters $K_1$ and $M_s$  the kinetic equations depend only on two parameters, the reduced viscosity $\tilde \eta $ introduced in Eq.(\ref{e:eta_tilde}) and 
rewritten as 
\begin{equation}
\tilde \eta = \eta / \eta_{water} \, \cdot (r_{d}/r_m)^3 .
\end{equation}
and a scaled temperature factor, 
 \begin{equation}
  {\cal T}=T/r_{m}^3.
\end{equation}
In these equations  $r_m$ denotes the radius of the magnetic core of the NP and  $r_d$ its hydrodynamic radius.

These relations  show in particular that an increase of the magnetic core region reduces the temperature fluctuation while an increase of the hydrodynamic volume of the NP  has a similar effect as an increase of the viscosity (for fixed $r_m$). It also means, for instance, that for a given ratio $r_{d}/r_{m}$ the parameter $\cal T$ goes to zero for a large particle size $r_{m}$ so that $p_{av}$ %absorbed power 
becomes independent of $r_{m}$ and approaches the $T=0$ value.  For intermediate particle sizes, on the other hand, a complex dependence on $r_m$ and $b_0$ is expected.

\subsection{Small fields}
To elucidate these dependences  we first show in  Fig. {\ref{f:fig_6} data for $p_{av}$  obtained  in the low field region  for $b_0 = 0.1$ and for $b_0=0.4$ (inset of Fig. {\ref{f:fig_6}).
The horizontal broken lines show $p_{av}$ for $T=0$ which is independent of magnetic radius $r_m$  as was discussed before. For finite temperatures, however,  sharp resonance type absorption peaks are observed exceeding the $T=0$-values significantly.  The peaks for $T=100$ K (circles, red) move to larger values of $r_m$ when increasing the temperature to  $T=300$ K (squares, blue). For increasing  $r_m $ the absorption approaches the $T=0$-values as is expected from the general discussion.

The reason for these observations is a temperature driven switching of the magnetization leading to an increase of the absorption in a restricted $r_m$-region. For very small $r_m$ switching is very fast resulting in a nearly reversible hysteresis curve with negligible energy absorption. For larger values of $r_m$ N\'eel -type relaxation sets in leading to large absorption rates. For even higher values of $r_m$ this type of relaxation is blocked resulting in a strong reduction of the absorption which, for very large values of $r_m$, approaches the $T=0$ - value as expected. The shift of the resonance-type absorption curve to higher values of $r_m$ with increasing temperature confirms this picture.

When increasing the field amplitude from $b_0 = 0.1$ to $b_0=0.4$ (inset  of Fig. 6)  the resonance type peaks broaden considerably and show not only an increase of  the maximal absorption by a factor of more than $10$ but also a shift of its  position to higher values of $r_m$ with increasing value of the field amplitude.   
 \begin{figure}[t] %  figure placement: here, top, bottom, or page
  \centering
 \includegraphics[width=3.5in]{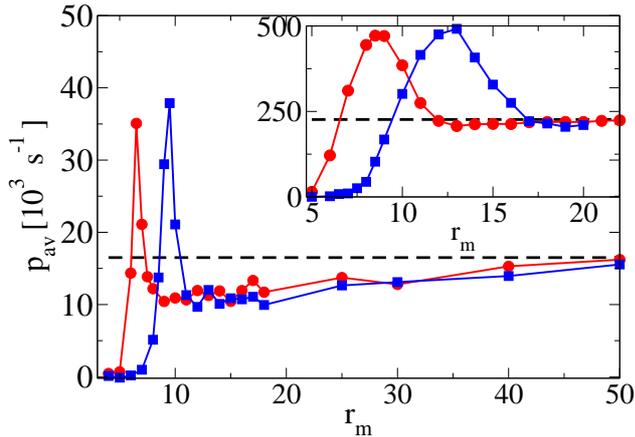}  \\[-0.1cm]
  \caption{(color online)  Averaged absorbed  power vs.  magnetic radius  $r_m$  for $b_0=0.1$. The inset shows results for $b_0=0.4$. For both graphs:  $T=0$ K (horizontal lines, black), $T=100$ K ( circles, red), and $T=300$ K (squares, blue);  $\tilde \eta=1.0$, $\alpha=0.2$,  $f\,=\,500$ kHz. }      
   \label{f:fig_6}
   \end{figure}     
   
For a deeper insight into these mechanism we calculated for $b_0=0.1$ the scalar product $\bf e \cdot \bf n$  in the stationary state. While for such small driving fields this quantity is very close to one  at $T=0$ independently  of $r_m$ it oscillates at finite temperatures  irregularly between the states  $\sim \pm 1$ on very small time scales. These temperature driven switching processes are very fast for small values of $r_m$ and slow down for increasing particle size up to a point where switching is blocked. This slowing down is shown in Fig. 7 for two values  of $r_m$ around the maximum of the absorption curve. The data are obtained in the stationary state  over a time interval  of length $\Delta \tau \sim 16 \, \tau_0$ where $\tau _0$ denotes the period of the driving field. The scalar product $\bf e \cdot \bf n$ is averaged over very small time intervals of length $\delta = \tau _0/100$,  $s = \langle {\bf e \cdot \bf n} \rangle _ \delta $, just in order to smoothen the data somewhat for drawing purposes. This quantity $s$  is plotted in Fig. 7 together with the driving field which is shown as wavy lines. It is seen that switching of the magnetization relative to the easy axis of the NP slows down significantly when increasing the size of the NP. 

The energy absorption is correlated with these temperature driven switching processes. For very small values of $r_m$ these processes are very fast and the response of the magnetization with respect to the driving field is almost reversible leading to small power absorption. Large power absorption is observed when the frequency of the external fields roughly matches the switching frequency which takes place in a narrow $r_m$ - interval.  A further  increase of the particle size  leads  to a blocking of the temperature assisted switching resulting in a sharp decrease of the power absorption towards its $T=0$ - value. 
 \begin{figure}[t] %  figure placement: here, top, bottom, or page
  \centering
 \includegraphics[width=3.5in]{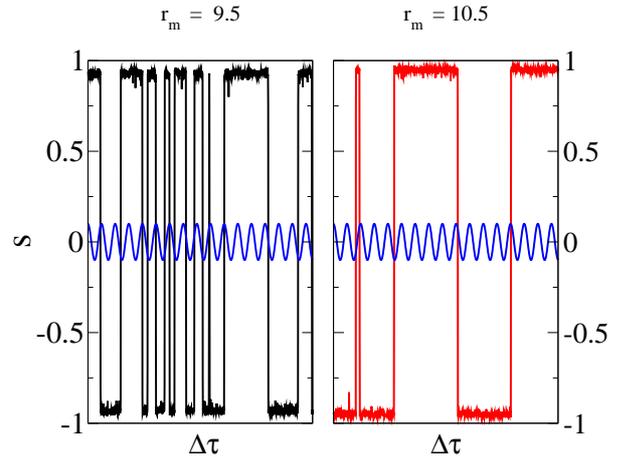}  \\[-0.1cm]
  \caption{(color online)  Averaged scalar product $s$  (see text for definition) for two values of $r_m$ vs. $\tau$ over time intervals of length $\Delta \tau$ together with driving field (wavy line, blue). 
 $b_0=0.1$,  $\tilde \eta=1.0$, $\alpha=0.2$,  $f\,=\,500$ kHz.}      
   \label{f:fig_7}
   \end{figure}     

\subsubsection{Immobilized NPs}

It is interesting to compare the results obtained with those for an ensemble of immobilized magnetic NPs. Out of the large number of theoretical papers  on the properties of such systems some of them have been devoted specifically to the problem of magnetic hyperthermia \cite{carr,hergt,raikher,usov_0}.  It is therefore of relevance to compare our results with those obtained for an ensemble of immobilized NPs. To this purpose we present numerical calculations on the power absorption  based on our kinetic equations. They are  reduced to the case of immobilized NPs using the same set of material parameters as before.  The ensemble  consists of 150 NPs with fixed anisotropy axis randomly distributed in space. In  Fig. 8  $p_{av}$ \,is shown for such an ensemble  of NPs for  $b_0=0.1$ and otherwise the same set of parameters as used in Fig. 6.

A comparison of the results  shows that the mobile NPs have a significant higher absorption rate  throughout. On the other hand the similarity in these curves suggests that in both cases the main absorption mechanism are similar.  This confirms our previous findings that a large power absorption is certainly due to magnetic switching processes of the magnetic moment relative to the NP. These switching processes are increasingly blocked for increasing $r_m$. This leads to a dramatic decrease of the absorption for immobile NPs with increasing $r_m$. For mobile NPs, on the other hand, freezing of the switching processes with increasing $r_m$ also takes place but  in this case $p_{av}$ levels out at the constant $T=0$-value the origin of which is a  significant absorption  due to Brownian rotational motion. 
 \begin{figure}[t] %  figure placement: here, top, bottom, or page
  \centering
 \includegraphics[width=3.5in]{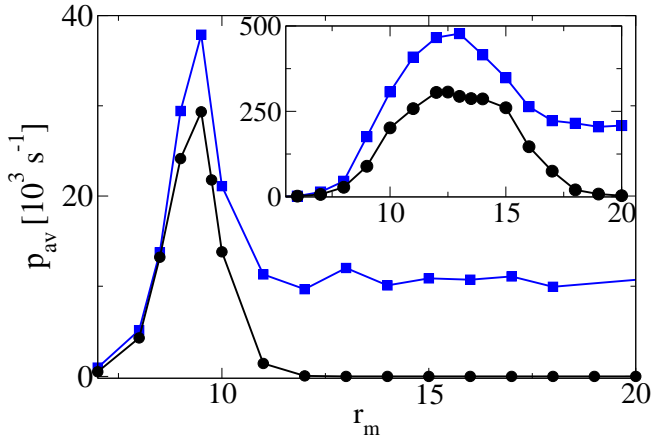}  \\[-0.1cm]
  \caption{(color online) Averaged absorbed power  for an ensemble of immobile NPs with randomly distributed anisotropy axis (circles, black) and for mobile NPs (squares, blue, same as in Fig. 6) for $b_0=0.1$. Inset shows results for  $b=0.4$.  $\tilde \eta=1.0$, $\alpha=0.2$, $f\,=\,500$ kHz.  }  
   \label{f:fig_8}
   \end{figure}

\subsubsection{Phenomenological approach}

In  \cite{rosen,nedelcu} a phenomenological approach was presented for heating magnetic fluids with alternating magnetic fields. It uses a simple relaxation equation for the magnetization of a NP with an effective relaxation rate $\tau_s$ consisting of two parts, Debye relaxation with relaxation time \cite{coffey_kal}
\begin{equation}
\tau_B=\frac{3 \eta V_d}{k_B T }
\end{equation} 
and N\'{e}el relaxation with relaxation time  \cite{brown, usad_2, coffey_kal}             
 \begin{equation}
\tau_N=\frac{1+\alpha ^2}{\alpha } \sqrt{k_B T/D} \, \, \pi/\omega_l \, \,\exp(D/k_B T).
\end{equation}
Arguing that Brownian and N\'{e}el processes take place in parallel the effective relaxation time $\tau_s$ in the linear response regime is supposed to be given by \cite{shliomis}
\begin{equation}
{\tau_s}^{-1} = \tau_B^{-1} + \tau_N ^{-1}
\label{e:tau}
\end{equation}  
leading to a power dissipation of the following form
\begin{equation}
 {\tilde p_{av}} \sim \frac{\tilde \omega \tau_s}{1+(\tilde \omega \tau_s)^2}.
 \label{e:pav}
\end{equation} 
An expression of this form has  widely been used, see for instance \cite{rosen, lahonian, nedelcu}.  This concept of an effective relaxation time, however, was criticized  to be not strictly justified  \cite{usov}. We note in passing that the energy absorption  derived for a set of rigid NPs in linear approximation \cite{carr,usov_0} also leads to an expression similar to that given in Eq. (\ref{e:pav}). 

 \begin{figure}[t] %  figure placement: here, top, bottom, or page
  \centering
 \includegraphics[width=3.5in]{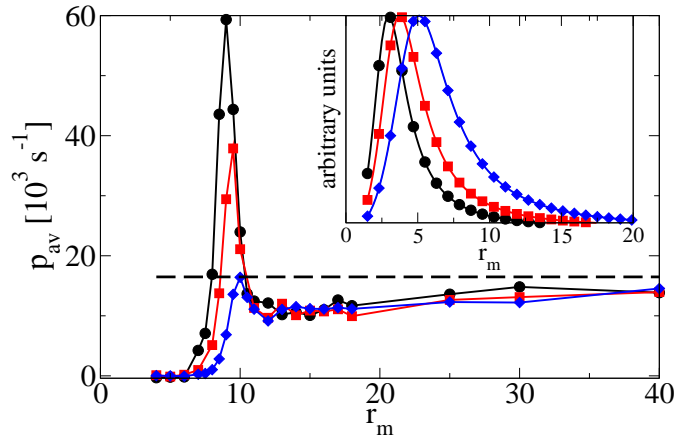}  \\[-0.1cm]
  \caption{(color online)  Averaged absorbed  power vs.  magnetic radius  $r_m$  for different values of frequency.  $f=1000$ kHz ( circles, black),  $f=500$ kHz ( squares, red), $f=100$ kHZ (diamonds, blue) for $b_0=0.1$, $T=\,300$ K, $\tilde \eta=1.0$, $\alpha=0.2$.  Horizontal broken line (black) shows 
 $p_{av}$ for $T=0$ K.
  The inset shows results obtained from Eq. 28.   }
   \label{f:fig_9}
   \end{figure}     
 
For a comparison of these heuristic results  with results obtained from our kinetic equations we show in Fig. 9  $p_{av}$  for different frequencies in the limit of small field amplitudes,  $b_0=0.1$. The inset of Fig. 9  shows  $\tilde p_{av}$ as given by Eq. (28) for the same set of parameters demonstrating that this simple approximation  leads indeed to a resonance-type absorption peak at a value of $r_m$ which, however, is only  about half of what we found in our calculations. 
Moreover, the absorption goes to zero for large $r_m$ contrary to our results where it approaches the constant $T=0$ -values as was discussed above. Additionally, frequency dependence and width of the absorption peaks exceed those found in our approach.  It is important to note that for this comparison no parameters are needed other than those we used in our numerical calculations.  
The qualitative agreement of these results with those from our more fundamental approach is remarkable, in particular because there are no adjustable parameters involved except for the over-all amplitude of $p_{av}$. It would be interesting to see under what assumptions and approximations our rather complicated equations can be reduced to such a simplified approach.

\subsection{Intermediate to high fields}

\begin{figure}[t] %  figure placement: here, top, bottom, or page
   \centering
 \includegraphics[width=3.5in]{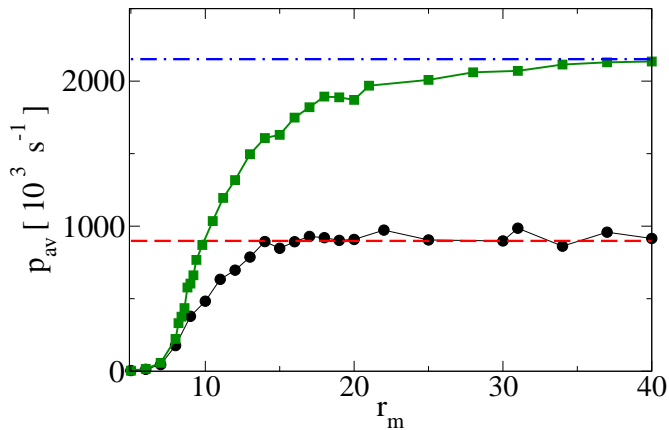}  \\[-0.1cm]
   \caption{(color online)  Averaged absorbed  power vs. magnetic radius $r_m$ for mobile NPs (squares, green) and for an ensemble of immobile NPs (circles, black). Dashed horizontal lines denote  the corresponding  $T=0$ -values.    
    $b_0=0.65$,  $T=300$ K, $\alpha = 0.2$,  $\tilde \eta = 2.0$, $f\,=\,1000$ kHz.  
  }      
   \label{f:fig_10}
   \end{figure}
\begin{figure}[t] %  figure placement: here, top, bottom, or page
   \centering
    \includegraphics[width=3.5in]{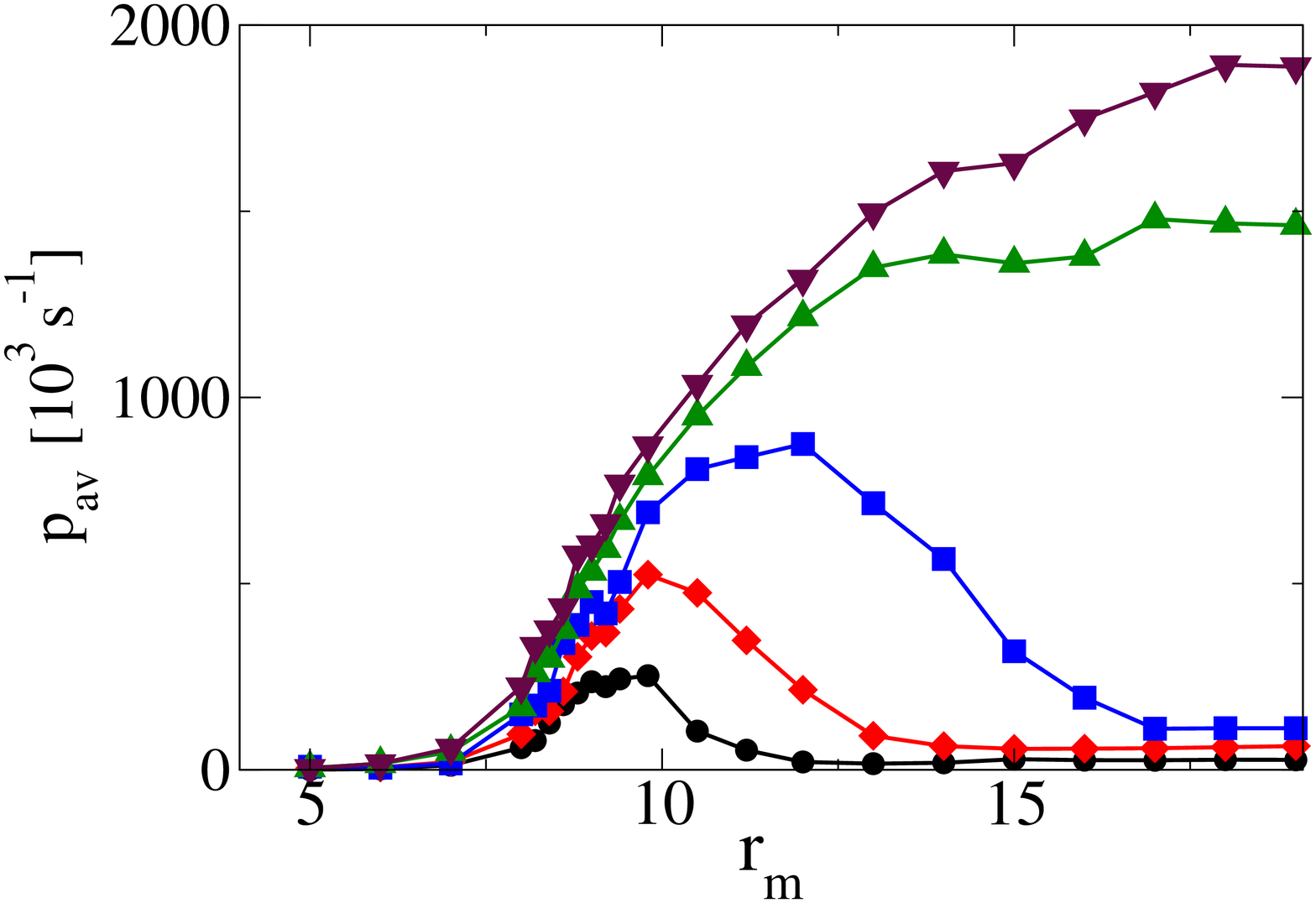}  \\[-0.1cm]
 \includegraphics[width=3.5in]{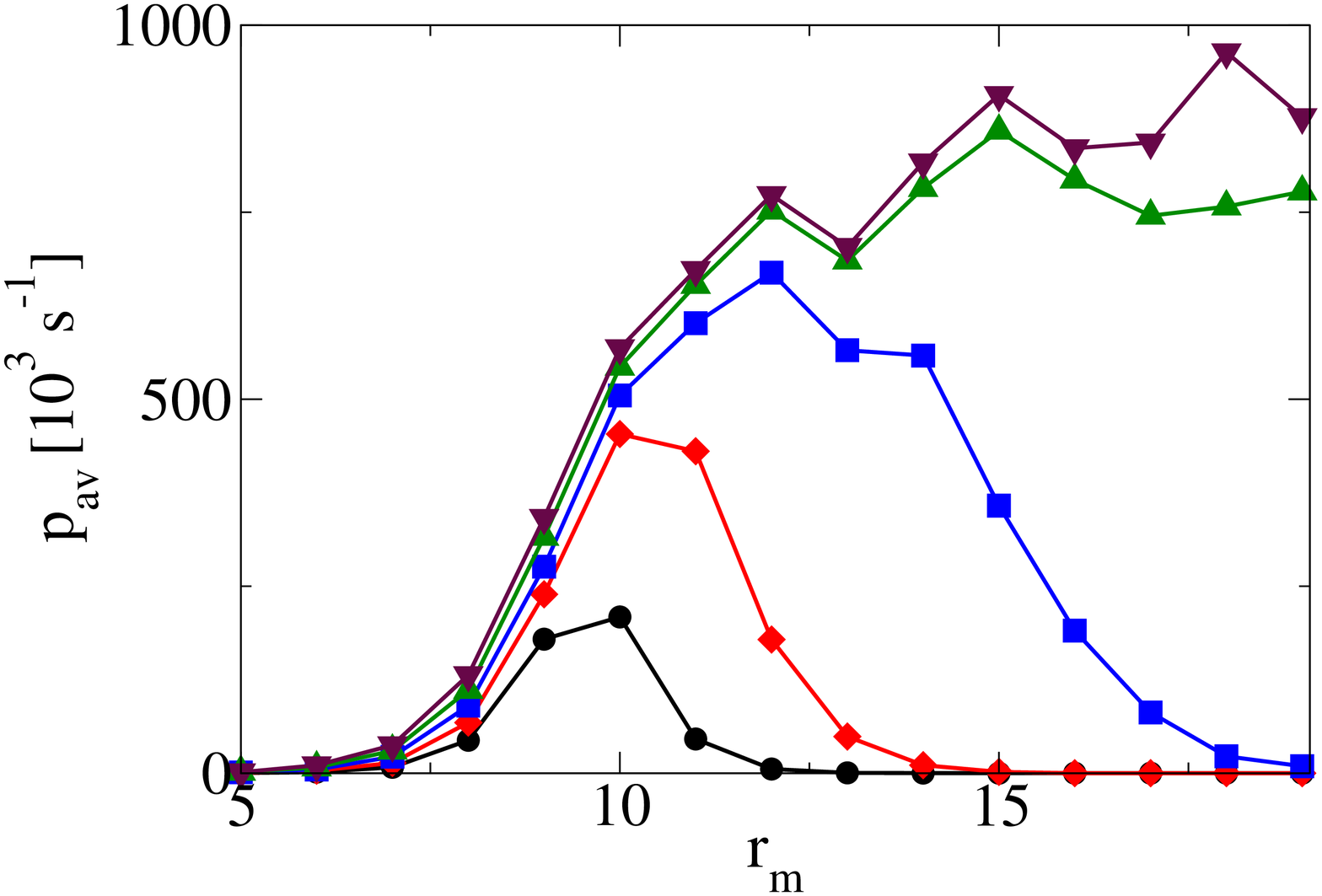}
   \caption{(color online) Upper panel:  Averaged absorbed  power vs.  $r_m$ for different field amplitudes. % 
 \newline
Lower panel:  Averaged absorbed  power vs. $r_m$ for an ensemble of NPs with fixed  randomly distributed anisotropy axis. \newline 
For both cases:   $b_0=0.65$: triangles down, maroon; $b_0=0.53$: triangles up, green; $b_0 = 0.4$:  squares, blue; $b_0=0.3$: diamonds, red; $b_0=0.2$: circles, black.  $T=300$ K, $\alpha = 0.2$,  $f\,=\,1000$ kHz.  }  
\label{f:fig_11}  
   \end{figure}

The broadening of the absorption peak and the shift of the position of its maximum to higher values of $r_m$  when increasing the field amplitude (see Fig. 6)  continues for fields up to about $b_0\sim 0.5$. For larger fields, on the other hand,  the peak structure disappears and a monotonic increase of $p_{av}$ as function of $r_m$ is observed for both mobile and immobile particles. This is  demonstrated in Fig. 10 for a field amplitude of $b_0 = 0.65$ and driving frequency $f=1000$ kHz.  The absorption increases as function of $r_m$  and approaches for large $r_m$ the $T=0$-values  from below.  The power absorption $p_{av}$ for an ensemble of mobile particles  is larger by more than a factor of two as compared to immobile particles. 

To be more specific, switching of the magnetization depends on temperature as expressed for instance in the N\'{e}el relaxation formula, Eq. (26), and on the value of the applied field, see for instance ref. \cite{coffey_kal}. As was discussed above, for small fields N\'{e}el relaxation is important leading to an extremely strong dependence of the relaxation time on the magnetic volume of the particle and on temperature leading in a restricted $r_m$ - interval even to an increase of the power absorption above its $T=0$ - value. For larger fields above $b_0 \sim 0.5$, on the other hand, switching  occurs already  without thermal fluctuations at $T=0$, c.f. Fig. 2. In this case thermal fluctuations reduce the absorbed power for all values of $r_m$ in contrast to low fields where they are essential for high power absorption. The largest reduction of $p_{av}$ occurs for small values of $r_m$ where thermal fluctuations are most important. 

More detailed results for $p_{av}$ in the region of small $r_m$ are presented in Fig. 11, both  for mobile and immobile particles.   The ensemble of immobilized magnetic NPs  consists of about 150 NPs with fixed anisotropy axis randomly distributed in space.  All the other parameters are the same in both parts of Fig. 11. 

The results show that the mobile NPs have a significant higher absorption rate  throughout. On the other hand the similarity of these curves again suggests that in both cases the main absorption mechanism are similar.  This confirms our previous findings that a large power absorption is certainly due to magnetic switching processes of the magnetic moment relative to the NP while viscous losses are relatively small except in those situations where magnetic switching processes are blocked so that viscous losses are the only sources for power absorption. 
   
The most interesting result being relevant for applications in hyperthermia, however,  is the development of a very broad maximum of $p_{av}$  as function of particle size already for moderate field amplitudes  of around $b_{0} \sim 0.4$. The maximal value of $p_{av}$  for $b_0=0.4$ is reached for $r_m \sim 12$ nm and exceeds the $T=0$-value by many orders of magnitude. Thus this temperature driven resonance type behavior has the important practical consequence that large absorption rates are already obtained for relatively small fields and for values of $r_m$ out of a broad interval. These findings are of particular interest because in applications a distribution in particle sizes is expected so that absorption rates which are not too sensitive to particle sizes are advantageous. 
On the other hand, for fields above the threshold region around $b_0 \sim 0.5$ a monotonous increase of  $p_{av}$ as function of $r_m$ is observed so that NPs with large magnetic radius $r_m$ are advantageous for power absorption.

\section{Conclusion}
In the present paper we proposed a set of equations for the dynamics of magnetic nano particles dissolved in a liquid. These equations are constructed on the basis of the LLG equation for classical spins together with  a classical equation of motion for a particle in a viscous medium taking the rotational degrees of freedom into account. The obtained kinetic equations are such that physically necessary conservation laws  are fulfilled in contrast to other published work. 

Numerical calculations reveal clearly that without thermal fluctuation two distinct regions can be identified  in a stationary state: for low amplitudes of the driving field the easy axis of the NP orients perpendicular to the field with a magnetization nearly fixed with respect to the easy axis of the NP. For larger fields the NP orients in such a way that its easy axis is nearly parallel to the direction of the driving field while its magnetization switches relative to the NP along its easy axis. A very sharp transition between these regions is observed at zero temperature.

The energy absorption calculated at finite temperatures increases considerably in the low field region and especially when going from the low field  to the high field region. Sharp resonance type absorption peaks as function of $r_m$ are observed for small field amplitudes which broaden and finally turn over to monotonic behavior when increasing the field amplitudes. These results contradict the conclusions drawn in \cite{usov} that in contrast to immobilized NPs the power absorption of NPs in a liquid increases monotonically with increasing particle diameter. Such a behavior is only observed for large enough field amplitudes. For smaller fields the choice for an optimal particle size leading to a large power absorption is more delicate because large absorption rates are only obtained in rather narrow $r_m$ - intervals. The illustrative numerical calculations at finite temperatures were performed for parameters relevant for hyperthermia and can help finding optimal parameters for an effective energy absorption by magnetic NPs.

The present work is motivated by applications in hyperthermia where magnetic NPs are likely to be free to rotate in a liquid. However, particles may also be immobilized at walls or inside biological cells \cite{fortin}. Results we obtained for ensembles of NPs with fixed anisotropy axis show that as far as power absorption is concerned the difference between these two cases is not as important as expected before \cite{usov}, at least not for parameters interesting for hyperthermia. 

In this contribution we only considered single nano particles  embedded in a viscous liquid. For an ensemble of interacting NPs a generalization of the present formalism is obvious although the numerical effort increases considerably. In this case not only the interaction of the particles has to be taken into account but also translational degrees of freedom because the particles will attract or repel each other affecting these degrees of freedom. Important in this context is the dipolar interaction which may lead to a clustering of particles depending on temperature \cite{hucht}.  Additionally, in an alternating field switching processes of the magnetization may result in changes of  the distance between particles leading to oscillations in the  particle positions which may generate high frequency oscillations in the liquid \cite{carr_sound}. These interesting questions arising from particle interactions are left for future research.

\end{document}